\documentclass[doublecol]{epl2}
\usepackage{aas_macros}
\usepackage{amsmath}
\usepackage{amssymb}
\usepackage{hubbleflow}
\usepackage{graphicx}
\usepackage{hyperref}
\usepackage{physics}
\usepackage[capitalize]{cleveref}
\usepackage{xcolor}

\hypersetup{
    bookmarks=true,         % show bookmarks bar?
    unicode=false,          % non-Latin characters in Acrobat’s bookmarks
    pdftoolbar=true,        % show Acrobat’s toolbar?
    pdfmenubar=true,        % show Acrobat’s menu?
    pdffitwindow=false,     % window fit to page when opened
    pdfstartview={FitH},    % fits the width of the page to the window
    pdfnewwindow=true,      % links in new window
    linkcolor=red,          % color of internal links
    citecolor=cyan,        % color of links to bibliography
    filecolor=magenta,      % color of file links
    urlcolor=blue,           % color of external links
    linktocpage=true
}

\newlength{\figw}
\newlength{\figh}
\newcommand{\efolds}{$e$-folds}

\newcommand{\umod}{\mathrm{mod}}
\newcommand{\usr}{\mathrm{sr}}
\newcommand{\up}{\mathrm{p}}

\newcommand{\calL}{\mathcal{L}}
\newcommand{\calE}{\mathcal{E}}
\newcommand{\calM}{\mathcal{M}}
\newcommand{\calI}{\mathcal{I}}
\newcommand{\calImod}{\calI_\umod}
\newcommand{\calIsr}{\calI_\usr}

\newcommand{\btheta}{\boldsymbol{\theta}}
\newcommand{\bthetapri}{\btheta_{\up}}
\newcommand{\bthetastd}{\btheta_{\us}}

\newcommand{\bdata}{\boldsymbol{D}}

\newcommand{\evid}[2]{\calE\negthinspace\left(#1|#2\right)}
\newcommand{\prior}[2]{\pi_{#1}\negthinspace\left(#2\right)}
\newcommand{\post}[2]{P\negthinspace\left(#1|#2\right)}
\newcommand{\prob}[1]{P\negthinspace\left(#1\right)}
\newcommand{\like}[2]{\calL\negthinspace\left(#1|#2\right)}

\newcommand{\nnow}{286}
\newcommand{\ASPIC}{\texttt{ASPIC}}
\newcommand{\CAMB}{\texttt{CAMB}}
\newcommand{\COSMOMC}{\texttt{COSMOMC}}
\newcommand{\POLYCHORD}{\texttt{PolyChord}}
\newcommand{\CAMSPEC}{\texttt{CamSpec}}

\title{Vanilla Inflation Predicts Negative Running}

\author{J\'er\^ome Martin\inst{1}\thanks{\email{jmartin@iap.fr}} \and Christophe Ringeval\inst{2,1}\thanks{\email{christophe.ringeval@uclouvain.be}} \and Vincent Vennin\inst{3,1}\thanks{\email{vincent.vennin@ens.fr}}}
\shortauthor{J\'er\^ome Martin, Christophe Ringeval and Vincent Vennin}

\institute{
  \inst{1} Institut d'Astrophysique de Paris, 98bis boulevard Arago,
  75014 Paris, France \\
  \inst{2} Cosmology, Universe and Relativity at Louvain (CURL),
  Institute of Mathematics and Physics, University of Louvain, 2 Chemin
  du Cyclotron, 1348 Louvain-la-Neuve, Belgium \\
  \inst{3} Laboratoire de Physique de
  l'\'Ecole Normale Sup\'erieure, ENS, CNRS, Universit\'e PSL,
  Sorbonne Universit\'e, Universit\'e Paris Cit\'e, 75005 Paris,
  France
}

\abstract{
  We show that the simplest, and currently favoured,
  theoretical realizations of cosmic inflation yield a sharp
  prediction for the running of the spectral index $\alphaS$. Using
  latest cosmological data, we compute its marginalized posterior
  probability distribution over the space of nearly $300$ models of
  single-field slow-roll inflation. The most probable value is
  $\alphaS=-6.3\times10^{-4}$, lying within the $98\%$ credible
  interval $-1.8 \times 10^{-3}<\alphaS< -9.1 \times 10^{-5}$. Within
  the landscape of all the proposed slow-roll inflationary models,
  positive values for the running are therefore disfavoured at more
  than three-sigma.}

\begin{document}

\maketitle

\section{Introduction}

Cosmic inflation is an epoch of accelerated expansion that took place
in the very early Universe, prior to the eras described by the hot
Big-Bang model. The main motivation for introducing this new phase of
evolution is that the resulting framework becomes free of the various
puzzles that plague the standard cosmological
model~\cite{Starobinsky:1980te, Guth:1980zm, Linde:1981mu,
  Mukhanov:1981xt}. Inflation comes in different incarnations but, at
least for the moment, the simplest realization, where inflation is
driven by a single scalar field with a minimal kinetic term (the
so-called inflaton field $\phi$), is sufficient to explain the
cosmological data. The mere existence of inflation addresses the
smallness of the spatial curvature today, the Gaussian statistics of
the Cosmic Microwave Background (CMB) anisotropies, the adiabaticity
of the initial conditions, and the presence of a nearly
scale-invariant primordial power spectrum $\calPz$ for curvature
perturbations. This last property has a special status as it was a
prediction of inflation (in contrast to a postdiction), which was
recently verified at a significant statistical level. Deviations from
scale-invariance are encoded in the value of the so-called spectral
index, $\nS$, evaluated at a pivot wavenumber $\kstar$
\begin{equation}
  \nS - 1 \equiv \eval{\dv{\ln\left[\calPz(k)\right]}{\ln
      k}}_{k=\kstar}.
\label{eq:nsdef}
\end{equation}
The latest cosmological data give, at $\kstar=0.05\,\Mpc^{-1}$, $\nS =
0.9649 \pm 0.004$ at $68\%$ confidence
limit~\cite{Planck:2018jri}. Because $\nS<1$, the primordial power
spectrum $\calPz(k)$ is slightly red, i.e., there is slightly more
power at low wavenumbers $k$ than at high wavenumbers.

\begin{figure}
  \begin{center}
    \includegraphics[width=\figw]{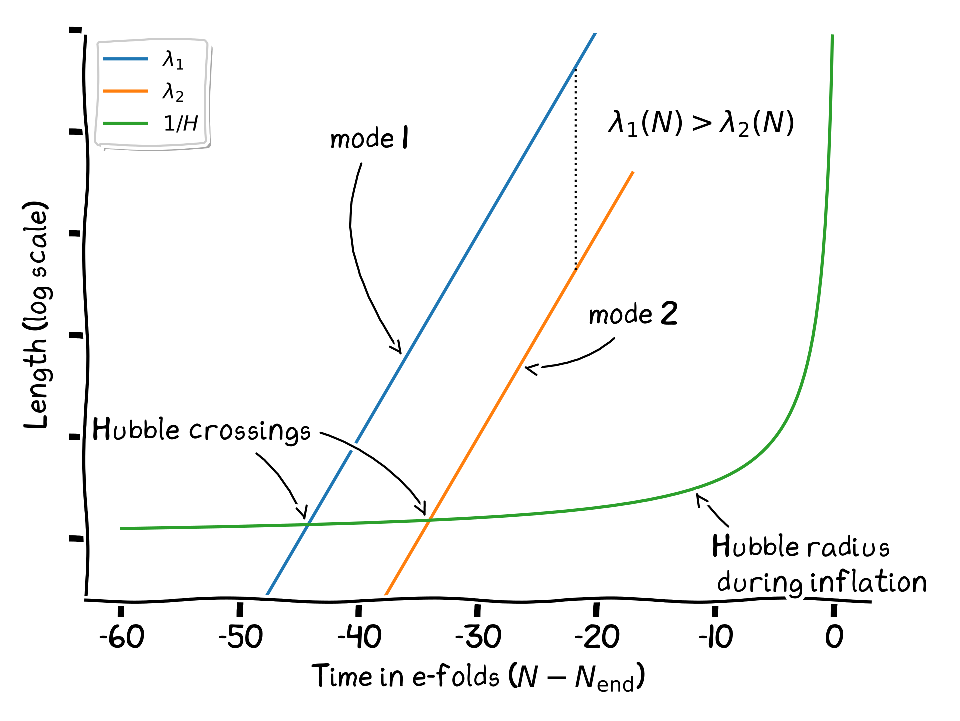}
    \caption{Evolution of two cosmological perturbations, of
      wavelengths $\lambda_1$ and $\lambda_2$, during cosmic
      inflation. Their amplitude is proportional to
      $H/\sqrt{\eps{1}}$, evaluated at Hubble crossing, i.e., when the
      wavelength equals the distance to the event horizon. In such a
      simple scenario, one has $\nS \lesssim 1$ together with $\alphaS
      \lesssim 0$.}
    \label{fig:sketch}
  \end{center}
\end{figure}
There are different ways of understanding why the simplest
single-field slow-roll scenarios mentioned above are associated with
$\nS \lesssim 1$. Indeed, typically for this class of models, the
potential $V(\phi)$ of the inflaton field is monotonous and, as
inflation proceeds, it becomes increasingly steep. As a consequence, the
inflaton accelerates and the kinetic energy grows compared to the
total energy so that, after a sufficient number of {\efolds}, denoted
$N=\ln a$, inflation comes to an end. This means that the Hubble-flow
functions, defined by~\cite{Schwarz:2001vv, Schwarz:2004tz}
\begin{equation}
\eps{i+1}(N) \equiv \dv{\ln \left|\eps{i}\right|}{N}\,, \quad
\eps{0}(N) = \dfrac{\Mp}{H}\,,
\label{eq:epsHdef}
\end{equation}
are typically increasing functions of time as inflation proceeds. For
instance, the physical process described above implies that
\begin{equation}
\eps{1} = \dfrac{1}{2\Mp^2}\left(\dv{\phi}{N}\right)^2,
\end{equation}
grows, which means that $\eps{2}>0$ thanks to \cref{eq:epsHdef}. At
leading order in the Hubble-flow functions, one can
show~\cite{Stewart:1993bc, Liddle:1994dx,
  Nakamura:1996da,Hoffman:2000ue, Gong:2001he, Leach:2002ar} that the
tensor-to-scalar ratio $r$ is given by $r=16\eps{1}$ and $\nS-1\simeq
-r/8-\eps{2}$. Given that primordial gravitational waves have not been
detected, current cosmological data only set the upper bound
$\log(\eps{1}) < -2.6$ ($95\%$)~\cite{Martin:2024qnn}, which
immediately implies that $\nS-1\simeq -\eps{2}$, that is to say a red
spectrum. From the value of $\nS \simeq 0.9649$ reported earlier, one
has $\eps{2} \simeq 0.035$.

Another way to describe the same mechanism in simple terms is to
remark that the amplitude of the power spectrum is given by $\calPz(k)
\propto H^2/(\Mp^2\eps{1})$, evaluated when the mode $k$ crosses out
the Hubble radius. When inflation proceeds, $H$ slightly decreases
(since $H^2\propto \rho$ can only decrease in an expanding universe)
while $\epsilon_1$ increases (if inflation, corresponding to
$\eps{1}<1$, ends by slow-roll violation when $\eps{1}$ reaches
unity). As a result, at late times, the amplitude of the mode
$\lambda_2 < \lambda_1$ is slightly lower than the one of $\lambda_1$,
see \cref{fig:sketch}. This implies that $\calPz(k_2) < \calPz(k_1)$
with $k_2 > k_1$: this is a red tilt.

A red tilt is thus typically expected if inflation is described by
vanilla models. In fact, the previous reasoning can be pushed one step
further. Indeed, in the same context, the running of the spectral
index is given by
\begin{equation}
  \alphaS  \equiv \eval{\dv[2]{\ln\left[\calPz(k)\right]}{(\ln
      k)}}_{k=\kstar}  \simeq -\eps{2}\left(2 \eps{1} + \eps{3}\right),
\label{eq:alphas}
\end{equation}
and, therefore, a positive third Hubble-flow function $\eps{3}>0$
implies a negative running. As such, one should expect the simplest
single-field scenarios to exhibit both $\nS \lesssim 1$ and $\alphaS
\lesssim 0$. It is important to notice that inflationary models having
$\alphaS > 0$ do exist, in the same manner as there are models with
$\nS \ge 1$. The main point here is that, within the aforementioned
simplest picture, there must be a strong correlation between the value
of $\nS$ and $\alphaS$. Because the current data strongly constrain
$\nS<1$, one should be able to have a definite prediction for
$\alphaS<0$, if the simplest scenarios were correct.

In the following, we quantitatively implement this idea and derive the
posterior probability distribution for $\alphaS$ given all the
explicit theoretical realizations of slow-roll single-field inflation
and using current cosmological data from both CMB and Baryonic
Acoustic Oscillations (BAO).

\section{Method}

We consider the space $\calImod$ of the single-field slow-roll
inflationary models studied in Ref.~\cite{EIoriginal}, which amounts
to $\nnow$ different scenarios. Each scenario, say $\calM_i$, comes
with some primordial model parameters $\bthetapri$ and their
associated prior
$\prior{i}{\bthetapri}=\post{\bthetapri}{\calM_i}$. The set
$\{\bthetapri\}$ is made of the parameters entering the shape of the
potential $V(\phi)$, the parameters describing the reheating era
after inflation, and any other parameters specific to the scenario
$\calM_i$ under scrutiny. A detailed description of each of these
models, parameters, and priors, can be found in
Refs.~\cite{EIoriginal, Martin:2013nzq, Martin:2024qnn}. Moreover, we
assume that, after the reheating, the universe evolves according to
the $\Lambda$CDM model, which is described by the standard
cosmological parameters, say $\bthetastd$. Given some cosmological
data sets $\bdata$, and a likelihood functional
$\like{\bdata}{\bthetapri,\bthetastd,\calM_i}$, one can determine the
posterior probability distribution of all parameters
\begin{equation}
\post{\bthetapri,\bthetastd}{\bdata,\calM_i} =
\dfrac{\like{\bdata}{\bthetapri,\bthetastd,\calM_i}
  \prior{i}{\bthetapri}\prior{}{\bthetastd}}{\evid{\bdata}{\calM_i}} \,,
\label{eq:bayesth}
\end{equation}
where the normalization constant is the Bayesian evidence
\begin{equation}
\evid{\bdata}{\calM_i} = \int
\like{\bdata}{\bthetapri,\bthetastd,\calM_i}\prior{i}{\bthetapri} \prior{}{\bthetastd}
\dd{\bthetapri} \dd{\bthetastd}.
\label{eq:eviddef}
\end{equation}
The probability of model $\calM_i$ to explain the data is then given by
\begin{equation}
\post{\calM_i}{\bdata,\calImod} = \dfrac{\evid{\bdata}{\calM_i}
  \prior{}{\calM_i}}{\prob{\bdata}}\,,
\label{eq:probaM}
\end{equation}
where $\prob{\bdata}$ is a model-independent normalization factor and
$\prior{}{\calM_i}=\post{\calM_i}{\calImod}$ is the prior probability of
model $\calM_i$ within the space $\calImod$. In the following, we
consider non-committal priors, i.e., $\prior{}{\calM_i} = 1/\nnow$ for
all models.

For all $\calM_i$, the Hubble-flow functions $\eps{i}(\bthetapri)$ are
known, and, from \cref{eq:alphas}, one can derive the running of the
spectral index $\alphaS(\bthetapri)$\footnote{In practice, we have
used the public library {\ASPIC} to compute the Hubble-flow functions
and the running~\cite{2018ascl.soft06031M}.}. All the posteriors
$\post{\bthetapri}{\bdata,\calM_i}$ can be obtained by performing
separated data analysis of $\bdata$ given $\calM_i$, and, from these,
it is straightforward to derive the posteriors of $\alphaS$ given $\calM_i$,
\begin{equation}
\post{\alphaS}{\bdata,\calM_i} = \int \post{\bthetapri}{\bdata,\calM_i}
\delta\left[\alphaS - \alphaS(\bthetapri)\right] \dd{\bthetapri}.
\end{equation}
We can then marginalize over all models $\calM_i$ filling the space $\calImod$
to obtain
\begin{equation}
\post{\alphaS}{\bdata,\calImod} = \sum_{i}
\post{\alphaS}{\bdata,\calM_i} \post{\calM_i}{\bdata},
\label{eq:postalpha}
\end{equation}
where $\post{\calM_i}{\bdata}$ is given by \cref{eq:probaM}. Let us
notice that each posterior appears weighted by the probability of the associated model,
ensuring that the most disfavoured scenarios have almost
no effect on the final result. The posterior probability
distribution $\post{\alphaS}{\bdata,\calImod}$ is the quantity of
interest: it is quantifying how probable a given value of the running
is given both the data sets $\bdata$ and the space of all models
$\calImod$.

\section{Results}

\begin{figure}
  \begin{center}
    \includegraphics[width=\figw]{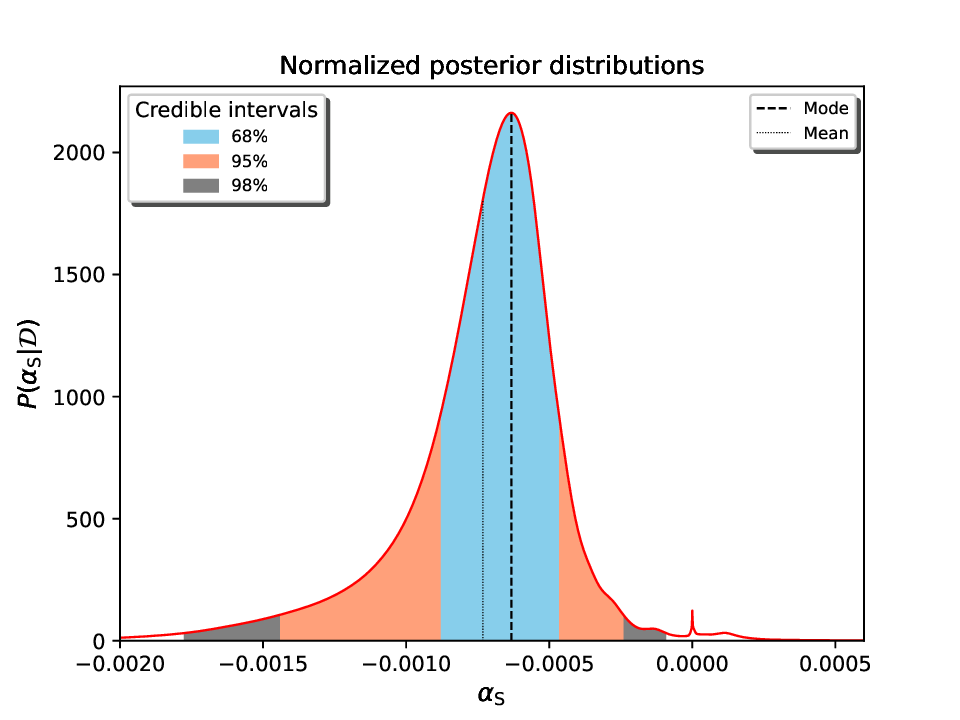}
    \caption{Posterior probability distribution for the running of the
      spectral index $\alphaS$ marginalized over $\nnow$ slow-roll
      inflationary models. Shaded regions show various annotated
      credible intervals comprising the mode and the mean
      value. Positive values of the running are disfavoured at more
      than three-sigma.}
    \label{fig:running}
  \end{center}
\end{figure}

The data sets $\bdata$ used are the 2020 post-legacy release of the
Planck satellite CMB data~\cite{Aghanim:2019ame,Efstathiou:2019mdh,
  Planck:2020olo, Tristram:2020wbi, Rosenberg:2022sdy}, complemented
with the BICEP/Keck array 2021 data~\cite{BICEP:2021xfz}, the South
Pole Telescope third generation measurements~\cite{SPT-3G:2021eoc} and
a compilation of BAO data from the Sloan Digital Sky Survey
IV~\cite{Dawson:2015wdb, SDSS:2017yll}. Data analysis has been done
using modified versions of the {\CAMB} and {\COSMOMC}
codes~\cite{Lewis:1999bs,Lewis:2002ah} to machine-learn an effective
likelihood~\cite{Ringeval:2013lea} which is fed to the {\POLYCHORD}
code ~\cite{2015MNRAS.453.4384H, Handley:2015fda}. The latter code is
used to perform fast nested sampling and evidence computation for the
$\nnow$ models. The details of this analysis can be found in
Ref.~\cite{Martin:2024qnn}.

Our main result is $\post{\alphaS}{\bdata,\calImod}$, which is
represented in \cref{fig:running}. The maximum and the
mean value of the running, a posteriori, are given by
\begin{equation}
  \label{eq:meanalpha}
    \left. \alphaS \right|_{\max} = -6.3\times 10^{-4}, \quad
    \mean{\alphaS} = -7.3 \times 10^{-4},
\end{equation}
respectively. They are both located within the following credible
intervals
\begin{equation}
  \begin{aligned}
-8.8 \times 10^{-4}   <  \alphaS < -4.7 \times 10^{-4} \quad (68\%),\\
-1.4 \times 10^{-3} < \alphaS < -2.4 \times 10^{-4} \quad (95\%), \\
-1.8 \times 10^{-3} < \alphaS < -9.1 \times 10^{-5} \quad (98\%),
  \end{aligned}
\label{eq:ci}
\end{equation}
which are represented as shaded regions in \cref{fig:running}. The
posterior $\post{\alphaS}{\bdata,\calImod}$ is almost vanishing in the
positive domain $\alphaS > 0$ and this implies that, indeed, all
single-field slow-roll models fitting well the current cosmological
data are associated with negative running.

Let us insist that \cref{eq:meanalpha} and Eqs.~(\ref{eq:ci}) can be
interpreted as predictions for $\alphaS$ within the space $\calImod$,
i.e., the space filled with all the simplest theoretical
implementations of single-field slow-roll inflation. If $\calImod$ is
truly representative of the actual way inflation occurred, future data
are going to match the figures quoted in \cref{eq:meanalpha} and
Eqs.~(\ref{eq:ci}). On the contrary, measuring positive values of the
running would imply that the simplest picture sketched in
\cref{fig:sketch} is incorrect and the evolution of the Hubble radius
has to be more complex than the one featured by vanilla models.

The data analysis we have presented is fully Bayesian, and, in order to
emphasize the importance of \cref{fig:running}, one also needs to
discuss its dependence with respect to the priors.

\section{Discussion}

\begin{figure}
  \begin{center}
    \includegraphics[width=\figw]{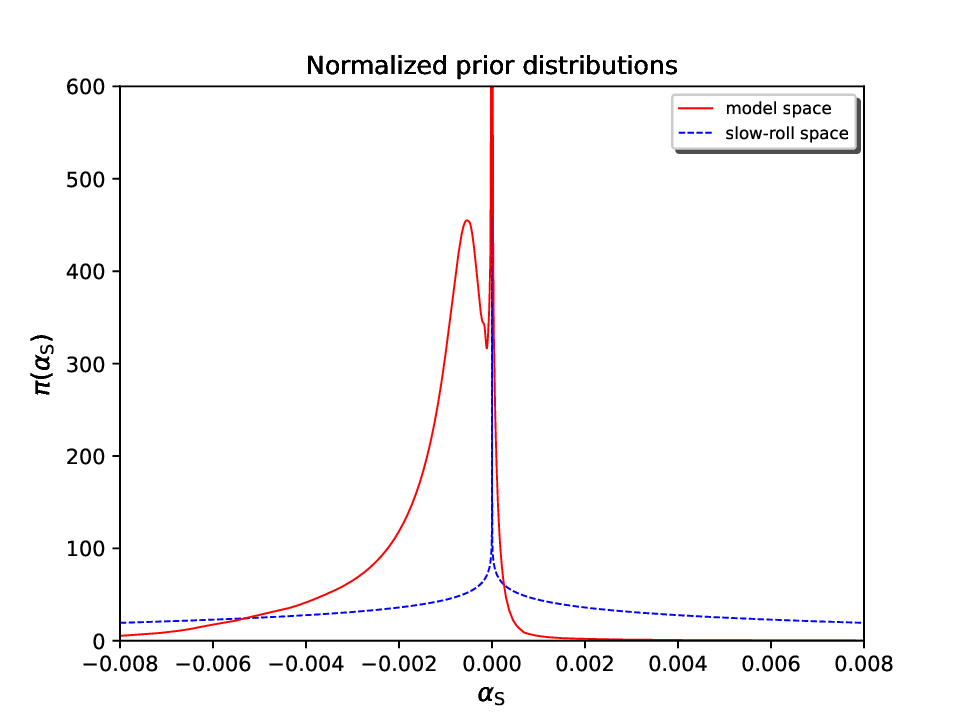}
    \caption{Prior probability distributions for the running of the
      spectral index $\alphaS$. The solid red curve is within
      $\calImod$, marginalized over $\nnow$ slow-roll inflationary
      models $\calM_i$ and assuming non-committal priors
      $\prior{}{\calM_i}=1/\nnow$. The blue dashed curve is the prior
      distribution assuming only slow-roll power spectra and standard
      priors on the Hubble-flow functions (see text).}
    \label{fig:compprior}
  \end{center}
\end{figure}

The prior probability distribution of the running is
$\prior{\umod}{\alphaS} = \post{\alphaS}{\calImod}$. From the previous
discussion, it is given by the weighted sum of all prior distributions
associated with the primordial parameters $\bthetapri$ of each models
$\calM_i$. One has
\begin{equation}
\prior{\umod}{\alphaS} = \sum_i \int \delta\left[\alphaS -
  \alphaS(\bthetapri)\right] \prior{i}{\bthetapri} \dd{\bthetapri} \prior{}{\calM_i}.
\end{equation}
This distribution has been plotted in \cref{fig:compprior} as a red
solid curve. It is not symmetric and negative values are clearly
preferred. Again, this is because the single-field slow-roll models
within $\calImod$ are mostly associated with smooth potentials and
behave as sketched in \cref{fig:sketch}. However, we should notice
that there is a tail in the positive region, signalling that some
models predict a positive running, and a sharp peak at vanishing
$\alphaS$. This peak comes from a subclass of inflationary models
having, for instance, a hilltop, or an inflection point, or very flat
regions in which the field gets trapped and inflation may end by
another mechanism than slow-roll violation. In this situation,
inflation is very close to pure de Sitter, the Hubble radius is very
close to constant and all Hubble-flow functions are very small. As a
consequence, these models are making definite predictions: $\nS\simeq
1$ and $\alphaS\simeq 0$, hence the peak. However, these de-Sitter
like scenarios have been disfavoured since the first measurements of a
red-tilted spectral index~\cite{2013ApJS..208...20B}, and, as evident
from \cref{fig:running}, only a small peak at $\alphaS=0$ remains in
the posterior $\post{\alphaS}{\bdata,\calImod}$. Comparing the width
of the prior in \cref{fig:compprior} to the posterior in
\cref{fig:running}, as well as the confidence intervals quoted in
\cref{eq:ci}, it is clear that our constraints, within $\calImod$, are
significantly driven by the data and are non-trivial.

\begin{figure}
  \begin{center}
    \includegraphics[width=\figw]{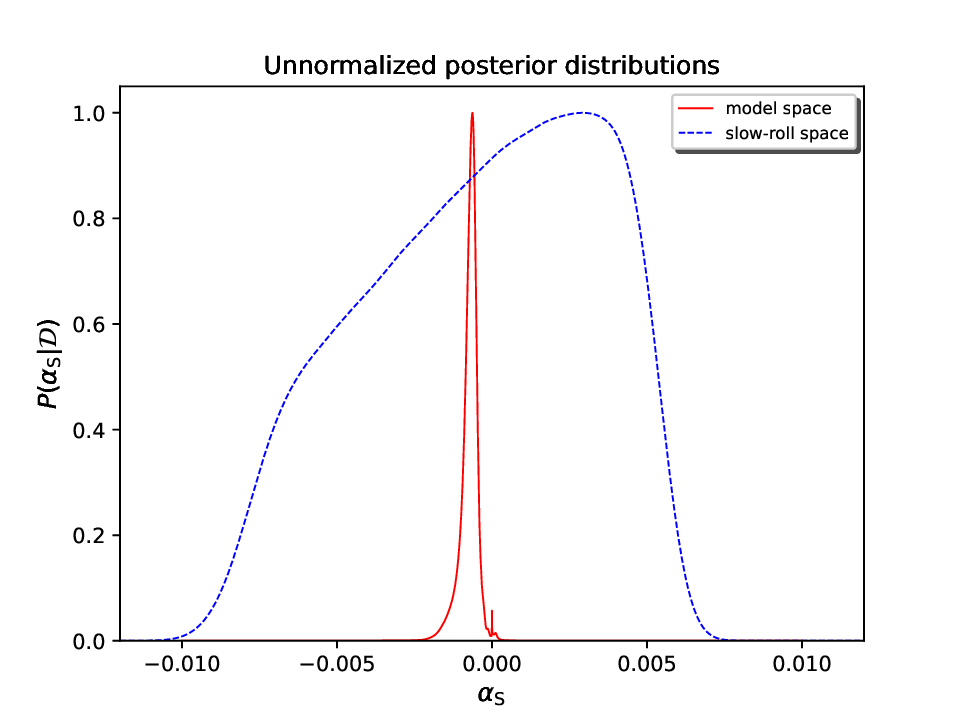}
    \caption{Comparison between the posteriors
      $\post{\alphaS}{\bdata,\calImod}$ (red solid line) and
      $\post{\alphaS}{\bdata,\calIsr}$ (blue dashed line). The former is
      associated with the space $\calImod$ of explicit theoretical
      realization of single-field slow-roll inflation whereas the
      latter, $\calIsr$, assumes only a Hubble-flow functional shape
      for the primordial power spectra. Explicit models, given the
      current data, are much more predictive.}
    \label{fig:comppost}
  \end{center}
\end{figure}

It is also informative to compare our results, applicable to
$\calImod$, to the ones that could be obtained starting from a much
wider space in the realm of slow-roll inflation. Let us define
$\calIsr$, the space of ``just slow-roll'' in which no explicit field
theoretical realization is assumed. The hypothesis space $\calIsr$
only assumes a slow-roll shape for the primordial power spectra, where
$\calPz(k)$ [and $\calPh(k)$ for tensors] are known functionals of the
Hubble-flow functions $\eps{i}$ (see, for instance,
Refs.~\cite{Martin:2013uma, Jimenez:2013xwa, Auclair:2022yxs} for
their explicit expression and derivation). From the motivated
priors\footnote{The first Hubble-flow function $\eps{1}$ is of unknown
order of magnitude whereas all $\abs{\eps{i}}$ should be smaller than
unity.}  $\log(\eps{1})\in[-5,-0.7]$, $\eps{2}\in[-0.2,0.2]$ and
$\eps{3}\in[-0.2,0.2]$, \cref{eq:alphas} allows us to compute
$\prior{\usr}{\alphaS} \equiv \post{\alphaS}{\calIsr}$, which is
plotted as a blue dashed curve in \cref{fig:compprior}. It is a
normalizable distribution sharply peaked at $\alphaS=0$, having very
long tails. It is symmetric and does not favour positive or negative
values of $\alphaS$. Using the data sets $\bdata$ described earlier,
we have derived the posterior distribution
$\post{\alphaS}{\bdata,\calIsr}$. It is represented in
\cref{fig:comppost} as a blue dashed curve. For convenience, we have
also reported the posterior $\post{\alphaS}{\bdata,\calImod}$ as a
solid red curve (same curve as in \cref{fig:running}). This figure
confirms that narrowing down the hypothesis space to $\calImod$, i.e.,
performing data analysis within explicit theoretical realizations of
single-field slow-roll inflation, significantly boosts the predictive
power of the cosmological data.

What would it take to change our conclusion? First, inflation could
have been realised by a \emph{vanilla model} not pertaining to $\calImod$.
If a large number of new models having totally different runnings,
fitting well the data\footnote{Otherwise their contribution to the
running posterior will be suppressed by their Bayesian evidence.},
were to be discovered, then our posterior distribution on the scalar
running would be substantially revised. Although this seems unlikely,
let us contemplate this possibility. The analysis of
Ref.~\cite{Martin:2024qnn} reveals that many of the single-field
models proposed after the release of the Planck 2013 data turn out to
be disfavoured (between 2013 and today, the fraction of rejected
models increased from 29 \% to 41 \%, although 112 additional models
were included in the recent analysis). This goes against the common
lore that theorists only propose models that are compatible with the
data available to them, and that the predictions of these models
reflect changes in the data rather than phenomenological typicality of
a class of theories. Instead, we argue that the models that have
already been proposed are representative of single-field constructions
in general, and that negative running does constitute a typical
prediction of vanilla inflation.

Second, one may want to work with \emph{model priors} differing from the
non-committal ones but, without a well-justified physical reason, this
would certainly be viewed as an awkward approach. In fact, if one were
to give more prior weight to models grounded in deeper theoretical
motivations than to potential functions proceeding from more
phenomenological parametrisations, one would exclude positive running
with even higher statistical confidence.

Third, inflation could have occurred in a \emph{non-vanilla
model}. This would come with additional signatures (entropic
perturbations, non-Gaussianities, \dots) that would significantly
modify Bayesian model comparison. At this stage however, these
signatures are constrained to be small by the data, hence the most
conservative and minimal assumption is that single-field slow-roll
inflation took place in the early universe. And this predicts negative
running. We conclude that it would be worth targeting future
observations~\cite{Hardwick:2018zry, Easther:2021eje} to reach the
figures quoted in \cref{eq:meanalpha} and Eqs.~(\ref{eq:ci}) to either
confirm it, or ruling it out.

\acknowledgements
We are indebted to Steven Gratton and Erik Rosenberg for having
provided us with their latest {\CAMSPEC} likelihood module for
{\COSMOMC}. This work is supported by the ESA Belgian Federal PRODEX
Grant $\mathrm{N^{\circ}} 4000143201$.

\bibliographystyle{JHEP}
\bibliography{biblio}

\end{document}